\documentclass[preprint,aps,amsfonts,amssymb,pdftex]{revtex4}

\usepackage{graphicx}
\usepackage{amsmath, amssymb, bm}   
\usepackage[dvips]{color}
\usepackage{natbib}
\usepackage{hyperref}
\usepackage[normalem]{ulem}  
\usepackage{xcolor}

\def\){\right)}
\def\({\left(}
\def\]{\right]}
\def\[{\left[}

\begin{document}

\title{The Four-Boson First-Excited State Near Two-Body Unitarity}

\author{
Feng Wu}
\affiliation{
Department of Physics, University of Arizona,
Tucson, AZ 85721, USA}

\author{
T. Frederico}
\affiliation{Instituto Tecnol\'ogico de
Aeron\'autica, 12.228-900 S\~ao Jos\'e dos Campos, SP,
Brazil.}

\author{
R. Higa}
\affiliation{Instituto de F\'isica, Universidade de S\~ao Paulo, R. do Mat\~ao Nr.1371, 05508-090, S\~ao Paulo, SP, Brazil}

\author{
U. van Kolck}
\affiliation{
Universit\'e Paris-Saclay, CNRS/IN2P3, IJCLab, 
91405 Orsay, France}
\affiliation{
Department of Physics, University of Arizona,
Tucson, AZ 85721, USA}

\begin{abstract}
Near two-body unitarity, the three-boson system is characterized by 
an approximate discrete scale invariance manifest in a geometric tower of bound states (the Efimov effect). In the absence of a strong four-body force, the four-boson system has two states associated with each Efimov state, one very nearly unstable, the other several times deeper. We study correlations between the excited- and ground-state properties, such as binding energies and radii, up to next-to-leading order in an effective field theory for short-range forces. We obtain the parameters in these correlations from similar correlations arising from existing precise calculations based on short-range potentials. We also derive correlations among excited-state properties that emerge from the proximity of the state to the break-up threshold into a boson and a three-boson bound state, using an effective field theory for ``halo'' states.
\end{abstract}

\date{\today}
\maketitle

\section{Introduction}
\label{sec:intro}

Weakly bound systems display a range of remarkable properties stemming from their large size compared to the range of interactions. Potential models with short-range interactions have been deployed effectively to describe weakly bound systems \cite{Frederico:2012xh}, but are not always in agreement. Short-Range (or Pionless in nuclear physics) Effective Field Theory (EFT) \cite{Hammer:2019poc} generalizes these models to highlight the universal behavior of few-body systems near the limit of two-body unitarity, as well as deviations from universality away from this limit. As such, it can be used to connect and differentiate various models. Here we deploy this framework to discuss the role of a four-body scale, the analog of the three-body scale that generates the Efimov effect \cite{Braaten:2004rn}.

Short-Range EFT consists of the most general short-range dynamics allowed by spacetime symmetries. A crucial ingredient is an {\it a priori} organization (``power counting'') of this dynamics according to their expected impact on observables, which is constrained by order-by-order renormalizability.
At leading order (LO), the unitary two-body system is at the scale-invariant non-trivial fixed point of the renormalization group (RG) \cite{Weinberg:1991um}. Renormalization of the three-body system requires a three-body force which lies on an RG limit cycle and reduces the symmetry to discrete scale invariance (DSI) \cite{Bedaque:1998kg,Bedaque:1998km}. This LO three-body force is sufficient to guarantee the proper renormalization of the four- \cite{Platter:2004he,Hammer:2006ct} and more- \cite{Bazak:2016wxm} body systems. As a consequence, properties of few-body systems near the unitarity limit are determined by a single three-body-force parameter $\Lambda_\star$, up to perturbative corrections \cite{Konig:2016utl,Konig:2019xxk}. However, this argument does not, by itself, prevent the appearance of a four-body force, and thus a four-body scale, already at LO.

Because at each order only a finite number of interactions enter, there exist correlations among observables which depend only on a limited set of parameters. Moreover, which parameters are relevant is a consequence of power counting. In this paper, we study how well several existing calculations conform to this framework, drawing in particular conclusions about the importance of a four-body force and the consequences of the extraordinarily large size of the first-excited four-body state.

In the absence of a four-body force at LO, the energies of all larger systems are correlated with the energy of the three-body ground state. The form taken by this correlation depends on the particle's statistics. To the extent they remain within the realm of applicability of the theory, bosonic systems form quantum-liquid drops \cite{Carlson:2017txq} and $N$-component fermionic systems tend to clusterize in $N$-body clusters \cite{Dawkins:2019vcr,Schafer:2020ivj}. 
In addition, spectra display DSI in the form of geometric towers of excited states. The first example was given by Efimov \cite{Efimov:1970zz} in the three-body system, where binding energies appear in a ratio $\simeq 515$. It has been found that each Efimov state is associated with two states in the four-body system \cite{Hammer:2006ct,Deltuva:2010xd}: one very near the three-body level and another $\simeq 4.6$ times more bound. Similar towers have been detected in systems with more particles \cite{vonStecher:2011zz,Gattobigio:2011ey,Gattobigio:2012tk,Kievsky:2014dua}.

Deviations from unitarity appear at next-to-leading order (NLO) in Short-Range EFT, in the form of a non-zero inverse two-body scattering length, a two-body correction linear in the energy (via the effective range) \cite{vanKolck:1998bw} and a four-body force needed for renormalization of the four-body system \cite{Bazak:2018qnu,Schafer:2022hzo}. These corrections improve the description of few-body systems away from unitarity \cite{Bazak:2018qnu}. For a discussion of the implications of DSI and its breaking, see Ref. \cite{vanKolck:2017jon}.

\begin{figure}[tb]
    \centering
    \includegraphics[width= 15cm]{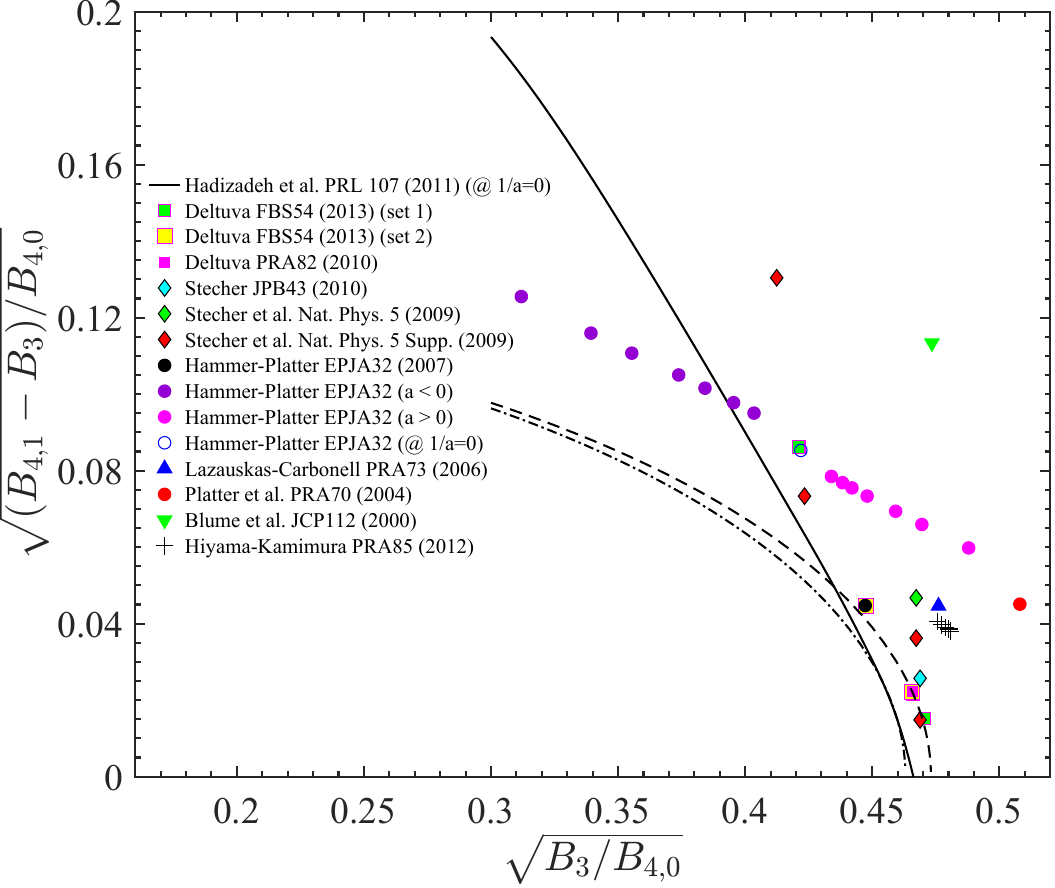}
    \caption{Correlation between four-body binding energies: the square root of the difference between the binding energies of the excited four-body state and the corresponding three-body state, namely $B_{4,\rm 1}-B_3$, as a function of the square root of the binding energy $B_3$ of the three-body state, both in units of the binding energy $B_{4,\rm 0}$ of the four-body ground state. Points stand for values from the literature: Refs.~\cite{Deltuva:2010xd,Deltuva:2012ig} (squares),
    Refs.~\cite{stechernature,stecherjpa} (diamonds), Refs.~\cite{Platter:2004he,Hammer:2006ct} (circles), Ref.~\cite{Lazauskas:2006} (up triangle),  Ref.~\cite{BlumeJCP00} (down triangle), and Ref.~\cite{Hiyama:2012cj} (crosses). The solid line is the zero-range calculation from Ref.~\cite{Hadizadeh:2011qj} at unitarity. The dashed and dot-dashed lines are obtained from Eqs.~\eqref{NLOcorrDeltuva} and \eqref{NLOcorrHadizadeh}, respectively.}
    \label{fig:1}
\end{figure}

The four-body force introduces a new scale and a new parameter in the correlation between excited- and ground-state four-body energies. This correlation was discovered in the context of short-range models by Hadizadeh {\it et al.} \cite{Hadizadeh:2011qj,Hadizadeh:2011ad}. In this approach, a zero-range two-body potential is supplemented by subtractions at three- and four-body levels, with two parameters. These three- and four-body parameters lead to a calculable correlation between the two four-body energies with respect to the three-body energy. It has been presented as a line on the plane of two binding energies --- the ``Hadizadeh plot'' \cite{Hadizadeh:2011qj}, see Fig. \ref{fig:1}. The correlation shown in this plot was found as a limit cycle computed using up to three tetramer states, which come to be bound as the four-body parameter is moved to large values with respect to the three-body parameter. This correlation is analogous to the one found for the energies of two successive Efimov trimer states (see Fig.~9 from Ref.~\cite{Frederico:2012xh}).
 
In Fig.~\ref{fig:1}, other results for the energy correlation are also summarized, which depend on the assumptions made in various calculations~\cite{Deltuva:2012ig,Deltuva:2010xd,stechernature,stecherjpa,Platter:2004he,Hammer:2006ct,Lazauskas:2006,BlumeJCP00,Hiyama:2012cj}. For example, Refs.~\cite{Deltuva:2010xd,Deltuva:2012ig} used only short-range two-body potentials, where the effects of a finite range were minimized by looking at states up in the tower. Results obtained from different states were close, but not indistinguishable, from the Hadizadeh line. They can be compared to results from two- and three-body Gaussian short-range potentials \cite{stechernature,stecherjpa} and realistic helium-helium interaction models~\cite{Lazauskas:2006,BlumeJCP00,Hiyama:2012cj}. Correlations for the energies of the atomic He trimers and tetramers for various potentials were presented in Ref.~\cite{Hiyama:2012cj}, where they were compared to results from Ref.~\cite{Platter:2004he}. LO calculations in Short-Range EFT~\cite{Platter:2004he,Hammer:2006ct} are also shown in Fig.~\ref{fig:1}.

Can we describe at least part of the correlation between four-body energies with Short-Range EFT at NLO? Answering this question should clarify the relation between approaches to few-body systems based on contact interactions and renormalization. Presently, there is no complete NLO calculation of the four-boson spectrum. In the pioneering Short-Range EFT calculation of Ref.~\cite{Hammer:2006ct}, results from finite values of the two-body scattering length have been used to extrapolate to the unitarity point. The large but finite scattering-length results are of course part of the full NLO correlation around the unitarity limit. Here we consider also correlations induced by the two-body effective range and the leading four-body force.

Correlations among other four-body quantities exist as well, again because at LO there is a single scale and three others appear at NLO. Some of these correlations can be related to energy correlations exploiting the ``halo'' nature of the excited state. For unknown reasons, this state is very close to the threshold for break-up into a three-body bound state and a particle. A lower-energy EFT, Halo/Cluster EFT \cite{Bertulani:2002sz}, can be deployed in which the three-body bound state is considered not a three-particle composite, but an ``elementary'' particle. This reduction from four degrees of freedom to two results in an additional expansion in the ratio between the characteristic sizes of the three-body and the associated excited four-body states. As an example, we apply this EFT to obtain information about correlations involving the four-body excited state's size. 

In addition to the four-body system being the simplest system sensitive to a four-body scale, the surprising proximity of the excited state to threshold and its extraordinarily large size make it very sensitive to corrections to DSI. The purpose of our paper is to exploit the consequences of the expected hierarchy of interactions through the correlations they generate at each order. A study of these correlations at LO and NLO forms the thread of this paper, which is organized as follows. In Sec. \ref{sec:LOCorr}, the energy correlations from scale symmetry breaking at LO and NLO are discussed for the excited four-body state. In Sec. \ref{sec:CorrPar} we use existing calculations to obtain the correlation parameters. We briefly discuss other correlations in Sec. \ref{sec:othercorr}. In Sec. \ref{sec:halo} we deploy Halo/Cluster EFT to estimate correlation parameters involving the excited-state size. Details about various radius definitions are relegated to Apps. \ref{sec:ric_rij} and \ref{rho1_HaloEFT}. Section \ref{sec:Conc} presents our conclusions.

\section{Energy Correlations}
\label{sec:LOCorr}

For nonrelativistic quantum systems governed by finite-range forces, the two-body unitarity limit can be reproduced by a nonderivative contact two-body force which contains no physical parameter. The presence of a single pole at zero energy in the two-body $T$ matrix determines the dependence of the two-body force on the regulator parameter, and the two-body system is scale invariant. As more particles (bosons or multi-component fermions) of the same species are considered, renormalization requires a nonderivative contact three-body force containing a dimensionful parameter $\Lambda_\star$ \cite{Bedaque:1998kg,Bedaque:1998km}. Only a discrete scale invariance remains, which implies that there is a tower of three-body states with a geometric series of binding energies \cite{Efimov:1970zz}. If there are two four-body states associated to the three-body ground state \cite{Hammer:2006ct,Deltuva:2010xd}, then DSI requires that there are two four-body states associated with each excited three-body state.

Short-Range EFT captures the low-energy manifestation of DSI, but can hold only up to a maximum momentum, which we denote by $M_{\rm hi}$, beyond which the details of the finite-range interaction cannot be neglected. The existence of $M_{\rm hi}$ reflects a partial breaking of DSI, which must appear in any physical theory. Short-Range EFT contains only semi-infinite towers of states on top of the ground states that, for a given number of particles, have the largest binding momenta below $M_{\rm hi}$. 

The single LO parameter $\Lambda_\star$ determines all low-energy observables. It can be determined from one three-body observable such as the three-body ground-state binding energy $B_3$. This observable may be obtained experimentally, or from the underlying theory, if that is known. (For a recent example of matching to an underlying theory with large, negative effective range, see Ref. \cite{Griesshammer:2023scn}.) All other observables calculated at LO are then correlated to $B_3$. In particular, we can write for the LO four-body ground and first-excited binding energies \cite{Platter:2004he,Hammer:2006ct}:
\begin{eqnarray}
B_{4,0}^{(0)} &=&\kappa_0 B_3,
\label{B400}
\\
B_{4,1}^{(0)} &=&\kappa_1 B_3.
\label{B410}
\end{eqnarray}
The universal numbers $\kappa_{0}>\kappa_{1}$ are obtained by solving the Schr\"odinger equation for a regularized contact potential followed by removal of regulator effects. They have been calculated precisely, $\kappa_{0}\simeq 4.6108$ and $\kappa_{1}\simeq 1.00228$ \cite{Deltuva:2010xd} (similar values with errors were given in Ref. \cite{Deltuva:2012ig}). Equation \eqref{B400} is a correlation known as the Tjon line, observed away from two-body unitarity with phenomenological potentials in nuclear \cite{Tjon:1975sme} and atomic systems \cite{Nakaichi:1978}. These potentials can be very different, but amount effectively to a three-body force with a single parameter once the two-body effect is reduced to a single contact interaction. Equation \eqref{B410} is a correlation that can be expressed in terms of the four-body ground state as 
\begin{equation}
B_{4,1}^{(0)} =\frac{\kappa_1}{\kappa_0} B_{4,0}^{(0)},
\label{B4ratio}
\end{equation}
where $\kappa_1/\kappa_0\simeq 0.21738$.

The same relations hold at LO in EFT when binding energies refer to any three-body state in the geometric tower and its associated four-body states.
By force, the ground and first-excited states are stable. However, states up the two four-body towers can decay into a particle plus a three-body state of lower energy. They are thus ``unstable bound states'' --- in the terminology of Ref. \cite{Badalian:1981xj} --- with a non-zero imaginary component of the energy, which is the negative of the half-width. The half-width also has to be proportional to $B_3$ of the associated Efimov trimer,
\begin{eqnarray}
\frac{\Gamma_{4,0}^{(0)}}{2} &=&\gamma_0 B_3
=\frac{\gamma_0}{\kappa_0} B_{4,0}^{(0)},
\label{Gamma400}
\\
\frac{\Gamma_{4,1}^{(0)}}{2} &=&\gamma_1 B_3
=\frac{\gamma_1}{\kappa_0} B_{4,0}^{(0)}.
\label{Gamma410}
\end{eqnarray}
The universal numbers $\gamma_{0,1}$ associated with the four-body states high in the tower have also been calculated in Ref. \cite{Deltuva:2010xd}: $\gamma_{0}\simeq 1.484 \cdot 10^{-2}$ and $\gamma_{1}\simeq 2.38\cdot 10^{-4}$ (again, similar values can be found in Ref. \cite{Deltuva:2012ig}).
These small numbers reflect the mismatch in size between a four-body state and a deeper three-body state. In fact, $\gamma_0$ is within a factor $\sim 1/3$ of the momentum ratio $\sim (22.7)^{-1}$ between adjacent three-body states. Thus, it is not surprising that as three-body states are removed below the physical three-body ground state by making them deeper and deeper, the widths of the two lowest physical four-body states vanish. 

These correlations represent the consequences of DSI. However, in a physical system DSI is broken by deviations from point interactions. When these deviations are small, they can be incorporated as perturbations in Short-Range EFT. New forms of the correlations arise which are distortions of the LO correlations. At NLO, the DSI-breaking interactions are included in first-order perturbation theory. The matrix elements of the perturbing interactions are calculated with respect to the LO wavefunctions, and thus are also universal and governed by DSI. We look here at the NLO modifications to the correlation \eqref{B4ratio}; analogous considerations to those below apply for the width \eqref{Gamma410}. 

One source of deviation from unitarity at NLO is a finite two-body $S$-wave scattering length. The towers of three-body states are frequently presented in the so-called ``Efimov plot'', where energies (determined by the strength of the LO non-derivative three-body contact force) are plotted as a function of the two-body scattering length (determined by the strength of a non-derivative two-body contact force). The plot can be generalized to include in addition to three-body energies also the four-body energies, which appear as two lines associated with every three-body line. 

At NLO in Short-Range EFT we need two more dimensions to incorporate the two additional NLO parameters: the $S$-wave effective range (determined by the strength of a two-derivative two-body contact force) and the four-body scale (determined by a non-derivative four-body contact force). The effects of the former can in principle be included in a four-body calculation just like the effect of a finite scattering length, but the stronger singularity of the interaction breaks RG invariance unless a four-body force is also present \cite{Bazak:2018qnu,Schafer:2022hzo}. The appearance of a four-body scale means a four-body datum is needed. This single four-body parameter leads to a correlation among four-body observables which cannot be determined in a model-independent way from fewer-body physics.

We first focus on the effect of the four-body scale by keeping the two-body system at unitarity. As a following step we incorporate finite two-body inverse scattering length and effective range. 
Note that we use a superscript $^{(1)}$ to denote results at NLO, which include both LO (with superscript $^{(0)}$) and NLO corrections.

\subsection{Four-body scale}

Retaining the two-body unitarity limit where the scattering length diverges, in addition to the vanishing of the effective range, we are limited to the plane of the three- and four-body parameters. The four-body energies form lines as the four-body parameter is varied. As the four-body force becomes more attractive, the two states should get deeper and more states become bound \cite{Frederico:2023fee}. As it gets less attractive, the lines must go towards zero energy. When they are close to vanishing the system is mostly an $S$-wave particle/three-body core system and a Halo EFT description must be possible. 

At NLO, when the four-body force is perturbative, four-body energies shift linearly with a four-body force parameter
that we will call $E_0^{(1)}$:
\begin{eqnarray}
B_{4,0}^{(1)} - B_{4,0}^{(0)} &=&\lambda_0(\Lambda) \, E_0^{(1)}(\Lambda),
\label{4bdeviation0}\\
B_{4,1}^{(1)} - B_{4,1}^{(0)} &=&\lambda_1(\Lambda) \, E_0^{(1)}(\Lambda),
\label{4bdeviation1}
\end{eqnarray}
where $\lambda_{0,1}(\Lambda)$ are regulator-dependent matrix elements of the four-body force. Their regulator dependence is compensated by that of $E_0^{(1)}(\Lambda)$, so that the binding energies become nearly cutoff independent at large cutoff values. These quantities depend entirely on LO physics and in principle do not vanish. 

One parameter is needed to fix $E_0^{(1)}(\Lambda)$, which we can take to be the most natural choice, $B_{4,0}^{(1)}$. Then $B_{4,1}^{(1)}$ can be calculated with only residual cutoff dependence. One consequence is that $\lambda_1(\Lambda)/\lambda_0(\Lambda)$ has a well-defined limit $\lambda_1/\lambda_0$ when the momentum cutoff increases arbitrarily. The asymptotic value $\lambda_1/\lambda_0$ cannot depend on $B_3$, since $\lambda_1/\lambda_0$ is dimensionless and there is no dimensionful parameter at LO other than $B_3$. As a consequence, $\lambda_1/\lambda_0$ should be the same for every three-body bound state. The asymptotic value of $\lambda_1/\lambda_0$ can be, but has not yet been, calculated. Until such calculation is performed we do not know even its sign, although one can expect $\lambda_1/\lambda_0>0$. But we do not know whether it is bigger or smaller than 1.

From Eqs. \eqref{4bdeviation0}, \eqref{4bdeviation1}, and \eqref{B4ratio},
\begin{equation}
\frac{B_{4,1}^{(1)}}{B_3} =
\kappa_1 -\frac{\lambda_1}{\lambda_0}\kappa_0
+ \frac{\lambda_1}{\lambda_0} \frac{B_{4,0}^{(1)}}{B_3},
\label{NLOcorr}
\end{equation}
where the first term on the right-hand side is the LO correlation \eqref{B410}, and the remaining terms represent the NLO correction 
rewritten in terms of the ground-state energy at NLO. This implies a universal, linear correlation between $B_{4,1}^{(1)}$ and $B_{4,0}^{(1)}$ in units of $B_3$, with a slope $\lambda_1/\lambda_0$ and an intercept determined by the same ratio and the known $\kappa_{0,1}$.

\subsection{Deviation from two-body unitarity}

In addition, we can consider small deviations from two-body unitarity when the scattering length satisfies $(a_2\sqrt{m B_3})^{-1}\ll 1$, where $m$ is the particle mass. This deviation can be considered in isolation, since it does not require a four-body force for renormalization. Deviation from unitarity can also arise from a natural-sized two-body effective range, $r_2\sqrt{m B_3}\ll 1$. This departure can be independent of the scattering length, but inclusion of the two-body two-derivative contact interaction that generates the effective range requires the presence of a four-body force.

Putting all the NLO effects together,
\begin{eqnarray}
\left(B_{4,0}^{(1)} - B_{4,0}^{(0)}\right)/B_3 &=&
\eta_0 \left(a_2\sqrt{m B_3}\right)^{-1}
+ \zeta_0 (\Lambda) \, r_2\sqrt{m B_3} 
+ \lambda_0(\Lambda) \tilde{E}_0^{(1)}(\Lambda)/B_3,
\label{total0}\\
\left(B_{4,1}^{(1)} - B_{4,1}^{(0)}\right)/B_3 &=&
\eta_1 \left(a_2\sqrt{m B_3}\right)^{-1}
+ \zeta_1 (\Lambda)\, r_2\sqrt{m B_3} 
+ \lambda_1(\Lambda) \tilde{E}_0^{(1)}(\Lambda)/B_3.
\label{total1}
\end{eqnarray}
In these expressions, the universal dimensionless numbers $\eta_{0,1}$ are cutoff independent. In contrast, $\zeta_{0,1}(\Lambda)$ are regulator dependent quantities and $\tilde{E}_{0}^{(1)}(\Lambda)$ is the four-body-force parameter in the presence of the effective range. Because this is first-order perturbation theory, $\lambda_{1}(\Lambda)/\lambda_{0}(\Lambda)$ must be the same as when $r_2=0$; in particular, it approaches a cutoff-independent $\lambda_{1}/\lambda_{0}$ for large cutoff. As is the case for $\lambda_{0,1}(\Lambda)$, $\eta_{0,1}$ and $\zeta_{0,1}(\Lambda)$ depend only on LO dynamics. 

Because $\eta_{0,1}$ are cutoff independent, we can consider the effects of the scattering length alone by setting $r_2=0$ and $\tilde{E}_0^{(1)}(\Lambda)=0$. The two-body scattering length induces correlated changes in $B_{4,1}$ and $B_{4,0}$. Following the same steps as before,
\begin{equation}
\frac{B_{4,1}^{(1)}}{B_3} =
\kappa_1-\frac{\eta_1}{\eta_0}\kappa_0
+ \frac{\eta_1}{\eta_0} \frac{B_{4,0}^{(1)}}{B_3} 
\qquad (r_2=0, \tilde{E}_0^{(1)}(\Lambda)=0)
\label{unicorr}
\end{equation}
describes the change in $B_{4,1}^{(1)}$ that accompanies a change in $B_{4,0}^{(1)}$ as a consequence of a variation in $a_2$. Again, since $\eta_1/\eta_0$ is dimensionless this correlation should hold throughout the three-body tower.

Once the range is included, the four-body force is compulsory. Now $B_{4,1}^{(1)}$ does not change in lockstep with $B_{4,0}^{(1)}$. If again we fit $\tilde{E}_0^{(1)}(\Lambda)$ to ensure $B_{4,0}^{(1)}$ is fixed in terms of experimental input, renormalization allows us to write 
\begin{eqnarray}
\frac{B_{4,1}^{(1)}}{B_3}&=&
\kappa_1 -\frac{\lambda_1}{\lambda_0}\kappa_0
+\frac{\lambda_1}{\lambda_0}\frac{B_{4,0}^{(1)}}{B_3}
+\left(\frac{\eta_1}{\eta_0}-\frac{\lambda_1}{\lambda_0}\right)
\eta_0 \left(a_2\sqrt{m B_3}\right)^{-1}
+ zr_2\sqrt{m B_3} \,,
\label{totalexcited}
\end{eqnarray}
where 
\begin{equation}
z \equiv \zeta_1 (\Lambda)- \zeta_0 (\Lambda)\frac{\lambda_1}{\lambda_0}
\label{z}
\end{equation}
is finite for large cutoff. This is the most general form of the energy correlation at NLO. The last two terms represent the effects of the two-body scattering length and effective range, and break DSI if we hold $a_2$ and $r_2$ fixed. For $r_2=0$, Eq.~\eqref{totalexcited} reduces to Eq.~\eqref{unicorr} for the scattering length that yields the full change in $B_{4,0}^{(1)}$, namely $(a_2 \sqrt{mB_3})^{-1}=(B_{4,0}^{(1)}/B_3 -\kappa_0)/\eta_0$. When instead $a_2^{-1}=0$, it reduces to Eq.~\eqref{NLOcorr}. The correlation expressed by Eq. \eqref{totalexcited} should hold as long as the corrections are perturbative, and apply across the three-body tower when all energies are rescaled. 

\section{Energy-Correlation Parameters}
\label{sec:CorrPar}

In Short-Range EFT, 
the correlation parameters in Eq. \eqref{totalexcited} are obtained from the expectation values of the NLO interactions with respect to the LO wavefunction, using a consistent regularization and renormalization scheme. Unfortunately, the wavefunction for the excited state is difficult to calculate reliably. However, some of the correlation parameters can be inferred from existing calculations, either directly in EFT or using potential models. In this section we interpret some of these calculations with Short-Range EFT.

\begin{figure}[tb]
    \centering
    \includegraphics[width= 15cm]{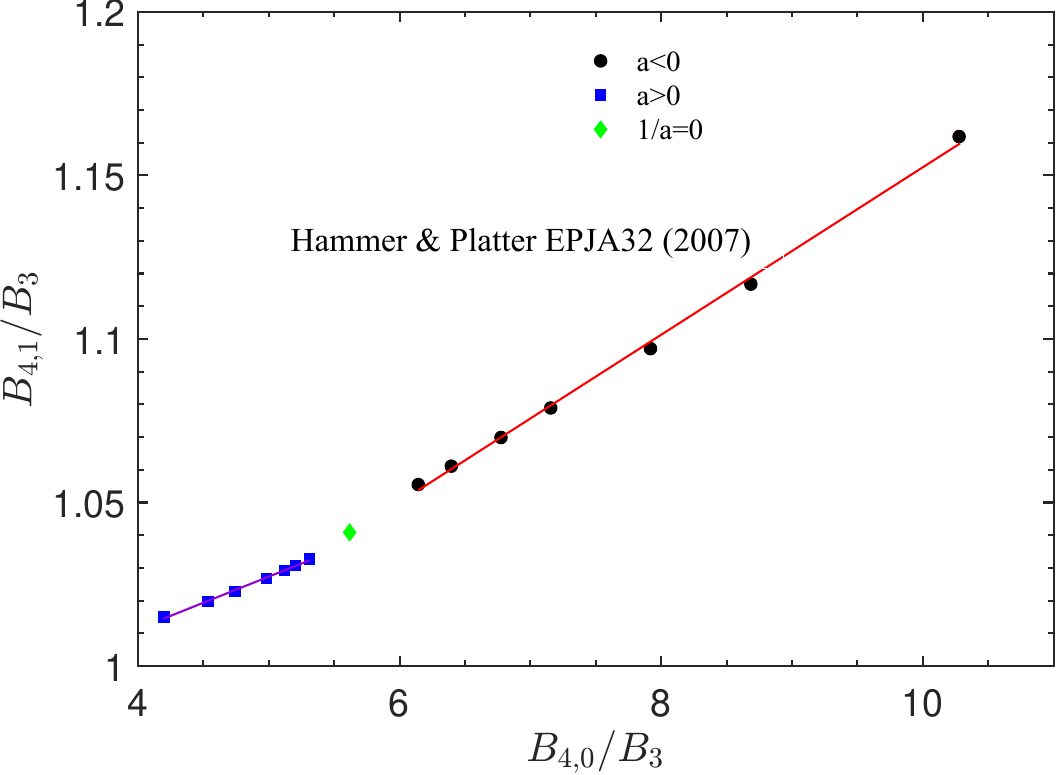}
    \caption{Energy of the four-body excited state $B_{4,1}$ as a function of the ground-state energy $B_{4,0}$, both in units of the energy $B_3$ of the associated three-body state, as the two-body scattering length is varied. The Short-Range EFT results of Ref. \cite{Hammer:2006ct} (as listed in Ref. \cite{Hadizadeh:2011qj}) for $a<0$ (full circles) and $a>0$ (full boxes) are fitted with Eq. \eqref{unicorr} (solid lines). The extrapolation to the unitarity limit is also given (full diamond).}
    \label{fig:HamPlat}
\end{figure}

For momentum $Q\sim a_2^{-1}$, the expansion around the unitary limit does not converge. Still, Short-Range EFT can be organized in such a way as to account for a finite $a_2^{-1}$ at LO --- that is, $(Qa_2)^{-1}$ terms are resummed --- in fact, most applications of Short-Range EFT do so \cite{Hammer:2019poc}. The two four-boson states were first calculated in this framework \cite{Hammer:2006ct}, and then $a_2^{-1}$ was varied. We collect these results (as given in Ref.~\cite{Hadizadeh:2011qj}) in Fig.~\ref{fig:HamPlat}. Fits with Eq. \eqref{unicorr}, which should be valid when $(a_2\sqrt{mB_3})^{-1}\ll 1$, yield
\begin{eqnarray}
\frac{B_{4,1}}{B_3}&\simeq& 0.896468+0.025601\; \frac{B_{4,0}}{B_3} 
\quad (a<0),
\\
&\simeq& 0.94742+0.0159823 \; \frac{B_{4,0}}{B_3} \quad (a>0).
\end{eqnarray}
The average slope is 
\begin{equation}
\frac{\eta_1}{\eta_0}\simeq 0.0207917.
\label{etaratio}
\end{equation}
The converged point $\kappa_0\simeq 5.61826$ and $\kappa_1\simeq 1.040858$, marked in Fig. \ref{fig:HamPlat}, is consistent with the average intercept 0.921944 since $\kappa_1-\kappa_0\eta_1/\eta_0\simeq 0.924045$. Presumably the calculation of Ref. \cite{Hammer:2006ct} differs from that of Ref. \cite{Deltuva:2010xd} because of range effects.

\begin{figure}[tb]
    \centering
    \includegraphics[width= 15cm]{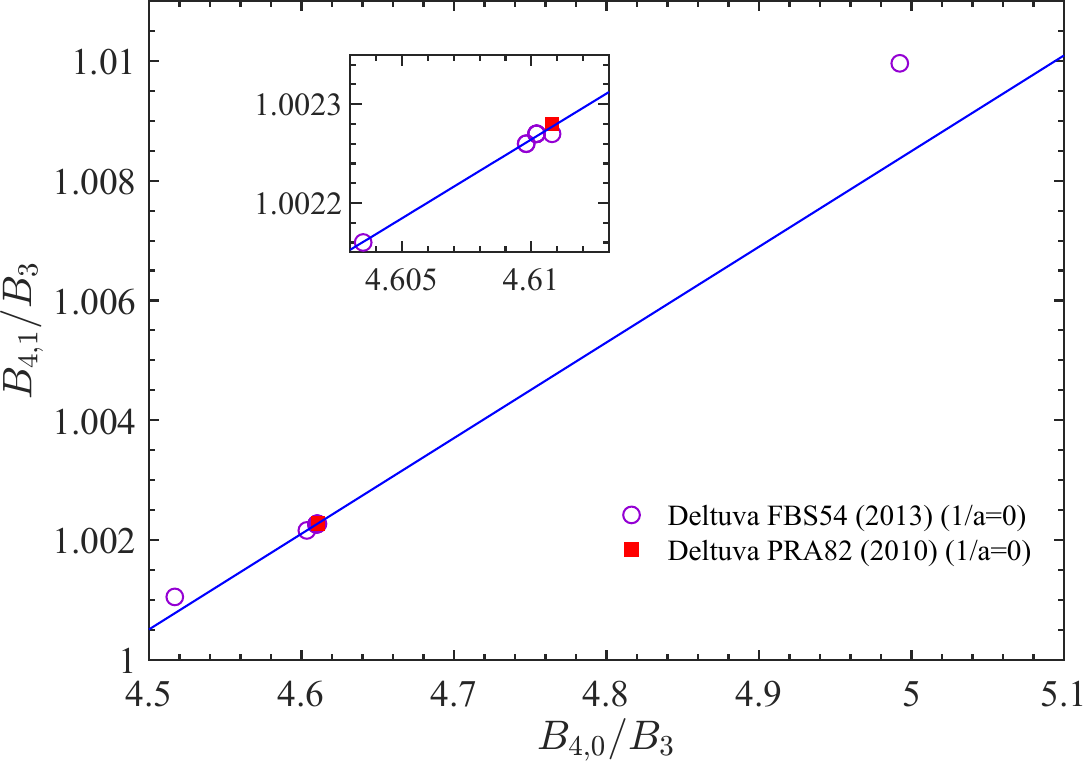}
    \caption{Energy of the four-body excited state $B_{4,1}$ as a function of the ground-state energy $B_{4,0}$,
    both in units of the energy $B_3$ of the associated three-body state, as one goes up in the Efimov tower. Results of excited ($n \geq 1$) states \cite{Deltuva:2012ig} for two separable two-body potentials (empty circles) are fitted with Eq. \eqref{NLOcorr} (solid line), neglecting the two extreme points, as shown in the inset. The converged result \cite{Deltuva:2010xd} is also indicated (full square).}
    \label{fig:Delt}
\end{figure}

The values of $\kappa_{0,1}$ in Ref. \cite{Deltuva:2010xd} are obtained from states sufficiently high in the Efimov tower when $a_2^{-1}=0$. Results for different states are shown in Table 2 of Ref.~\cite{Deltuva:2012ig}, starting from the ground state, for two separable two-body potentials (sets 1 and 2). Points for excited states ($n \geq 1$ states in Ref. \cite{Deltuva:2012ig}) are plotted in Fig.~\ref{fig:Delt}. They are approximately lined up, except for the $n=1$ points at the extremes where range effects are still important. We fit the other points (points $n=$2, 3, 4, and 5 for sets 1 and 2) using Eq. \eqref{totalexcited}, assuming that the last two terms on the right-hand side are absent, that is, using Eq. \eqref{NLOcorr}. We obtain
\begin{equation}
\frac{B_{4,1}^{(1)}}{B_3} \simeq 0.928624 + 0.0159741 \frac{B_{4,0}^{(1)}}{B_3}.
\label{NLOcorrDeltuva}
\end{equation}
Thus the slope gives one of the correlation parameters,
\begin{equation}
\frac{\lambda_1}{\lambda_0}\simeq 0.0159741.
\label{lambdaratio}
\end{equation}
The value for the intercept is consistent with the values for $\kappa_{0,1}$ \cite{Deltuva:2010xd} which, together with the slope \eqref{lambdaratio}, give $\kappa_1-\kappa_0\lambda_1/\lambda_0\simeq 0.928627$. The $n=1$ points of sets 1 and 2 fall above the line by $0.0003$ and $0.0016$, respectively. This is consistent with the last term on the right-hand side of Eq. \eqref{totalexcited} if $z>0$. For the other excited states, this term is expected to be smaller by factors of $\simeq 22.7$ and is thus negligible, justifying the use of Eq. \eqref{NLOcorr}.

Results with a four-body scale were obtained at and near the unitarity limit in Refs.~\cite{Hadizadeh:2011qj,Hadizadeh:2011ad}. In the four-boson Faddeev-Yakubovsky (FY) equations for the zero-range two-body interaction, the three- and four-body scales can be introduced by subtractions in the appropriate four-body Green's functions, in a way to keep the three-body binding fixed while the four-body energy is moved~\cite{Hadizadeh:2011qj}. As the four-body attraction increases additional four-body states aggregate below threshold and the behavior of four-body binding energies resembles~\cite{Frederico:2019bnm} the limit cycle found at three-body level, when continuous scale symmetry is broken to a discrete version. In a Hamiltonian framework, the dependence of four-body states on a four-body scale is translated into a four-body short-range interaction, which within EFT is introduced pertubatively at NLO~\cite{Bazak:2018qnu}. Therefore, a portion of the results of Refs.~\cite{Hadizadeh:2011qj,Hadizadeh:2011ad} should be described in terms of the correlations obtained in Short-Range EFT.

\begin{figure}[tb]
    \centering
    \includegraphics[width= 15cm]{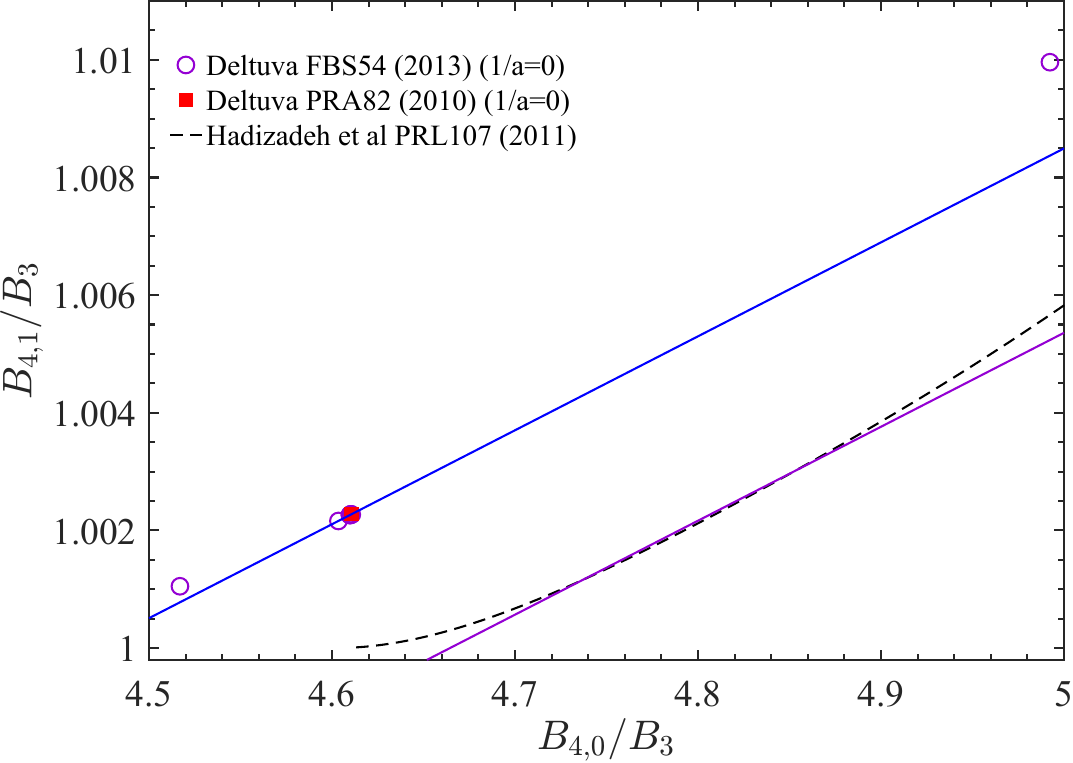}
    \caption{Energy of the four-body excited state $B_{4,1}$ as a function of the ground-state energy $B_{4,0}$, both in units of the energy $B_3$ of the associated three-body state, as a four-body scale is varied. Results \cite{Hadizadeh:2011qj} for a nonpertubative implementation of this scale (dashed line) are fitted with Eq. \eqref{NLOcorr} (solid line).}
    \label{fig:Hadi}
\end{figure}

In order to allow an EFT interpretation, we compare in Fig.~\ref{fig:Hadi} the FY calculations of Deltuva~\cite{Deltuva:2010xd,Deltuva:2012ig} and Hadizadeh {\it et al.}~\cite{Hadizadeh:2011qj} at unitarity. We construct a line parallel to Deltuva's results, that is, we use the slope \eqref{lambdaratio} and adjust the intercept, 
\begin{equation}
 \frac{B_{4,1}}{B_3}\simeq 0.9255 + 0.0159741\, \frac{B_{4,0}}{B_3}\, ,
\label{NLOcorrHadizadeh}
\end{equation}
in order to overlap partially with Hadizadeh {\it et al.}'s curve. The results from Hadizadeh {\it et al.} show more features than the linear dependence, which holds well within the interval $4.7\lesssim B_{4,0}/B_3\lesssim 4.9$. The intercept is slightly different from that obtained in Fig.~\ref{fig:Delt}. A possible origin for the disparity~\cite{DeltuvaPC} is that in the calculations of Ref.~\cite{Hadizadeh:2011qj} the FY components were truncated to the $S$ wave, while in Ref.~\cite{Deltuva:2012ig} higher partial waves were considered. It is remarkable that the slope is common to these two independent calculations, one without any four-body force and the other obtained with the introduction of a four-body scale in nonperturbative form. From the EFT perspective, both approaches are effectively introducing a four-body force, which leads to an energy correlation through $\lambda_1/\lambda_0$.

For another way to see the differences between Short-Range EFT and Hadizadeh {\it et al.}'s calculation, we can return to Fig. \ref{fig:1}. There, in addition to the results of existing calculations, we plot
\begin{equation}
\sqrt{\frac{B_{4,1}^{(1)}-B_3}{B_{4,0}^{(1)}}}=
\sqrt{\left(\kappa_1-\frac{\lambda_1}{\lambda_0}\kappa_0-1\right)
\left(\sqrt{\frac{B_3}{B_{4,0}^{(1)}}}\right)^2
+ \frac{\lambda_1}{\lambda_0}}
\label{NLOcorrprime}
\end{equation}
for $\lambda_1/\lambda_0$ given in Eq. \eqref{lambdaratio} and the two values for the intercept $\kappa_1-\kappa_0\lambda_1/\lambda_0$: 0.928624 (from Fig. \ref{fig:Delt}) and $0.9255$ (from Fig. \ref{fig:Hadi}). By construction these lines describe well the results from Refs.~\cite{Deltuva:2010xd, Deltuva:2012ig} and \cite{Hadizadeh:2011qj}, respectively, at small excited four-body energy.

Similarly to Deltuva's extreme points, Hadizadeh {\it et al.}'s curve in Fig.~\ref{fig:Hadi} shows deviations from a straight line, which are reflected in the gentler curvature and slope with which it approaches the $x$ axis in Fig.~\ref{fig:1}. These deviations are consistent with effective-range effects where $z>0$ as defined in Eq. \eqref{z}. However, there are no explicit range effects in Ref.~\cite{Hadizadeh:2011qj} and other explanations for this curvature are possible. On the left side of Fig.~\ref{fig:Hadi}, the excited state is becoming unstable, presumably turning into a virtual state of the particle+three-particle system. This is the mechanism by which a three-body Efimov state becomes unstable as the scattering length $a>0$ decreases (see, e.g., Ref.~\cite{Frederico:2012xh}). It is possible that such a critical point cannot be properly characterized by perturbative NLO corrections in Short-Range EFT. On the right side, the four-body force increases and we might be seeing the appearance of higher-order effects. Note that in the range of Fig.~\ref{fig:Hadi} the deviations from a straight line are small compared to the difference between the Hadizadeh {\it et al.}'s and Deltuva's curve, which presumably~\cite{DeltuvaPC} stems from the approximations made in Ref.~\cite{Hadizadeh:2011qj}.

These results suggest that there is no conflict between Short-Range EFT and the calculations of Hadizadeh {\it et al.} When the implicit four-body force of the latter is relatively weak, these calculations can be reproduced at NLO in Short-Range EFT with the correlation parameter \eqref{lambdaratio} and a small adjustment in $\kappa_1-\kappa_0\lambda_1/\lambda_0$. When the implicit four-body force is relatively strong, the Hadizadeh {\it et al.} calculations apply to a class of systems where a four-body scale would appear already at LO in Short-Range EFT, even though it is not required by renormalization.

We have thus obtained two of the energy-correlation parameters that appear in Eq. \eqref{totalexcited}: $\lambda_1/\lambda_0$, Eq. \eqref{lambdaratio}, and $\eta_1/\eta_0$, Eq. \eqref{etaratio}. Unfortunately we were not able to determine the remaining parameter, the $z$ of Eq. \eqref{z}.

\section{Other Correlations}
\label{sec:othercorr}

Similar correlations through $\Lambda_\star$ exist in Short-Range EFT for other four-body quantities.
As an example, we consider here mean-square radii $\langle r^2 \rangle_{4,0}^{(n)}$ and $\langle r^2 \rangle_{4,1}^{(n)}$, defined as the average squared distance of a particle from the center of mass, for the ground state and first-excited state at N$^n$LO. These radii, which we also denote by $\langle r_{iG}^2 \rangle$ ($G$ stands for center of gravity) in the following, are related to the average squared interparticle distance $\langle r_{ij}^2 \rangle$ in App. \ref{sec:ric_rij}.

The four-body states, apart from the lowest two physical states, are unstable bound states. This is reflected not only in their widths but also more generally in other quantities, such as radii, taking on complex values. Their imaginary parts are also correlated to other observables but, being tied to deeper states, are small. When the breakdown scale is accounted for, the Efimov tower is truncated at the ground state and the radii for the lowest two tetramer states become real as their widths vanish. For illustration, we focus here on the real part of the radius, $\text{Re}\langle r^2 \rangle$, which at LO are correlated to their corresponding binding energies through
\begin{eqnarray}
m\left(B_{4,0}^{(0)}-B_{3}\right) \text{Re} \langle r^2 \rangle_{4,0}^{(0)} 
&=& \rho_0 ,
\label{r400}
\\
m\left(B_{4,1}^{(0)}-B_{3}\right)\text{Re}\langle r^2 \rangle_{4,1}^{(0)}  &=& \rho_1.
\label{r410}
\end{eqnarray} 
The numbers $\rho_{0,1}$ obtained at large cutoff are universal as they are dimensionless and there is no other scale at LO than $\Lambda_\star$. 
Using the LO binding energy and point-charge radius (normalized by the charge) obtained for the ground state of the $^4$He nucleus at unitarity~\cite{Konig:2019xxk} with finite cutoffs, we find the universal number $\rho_0= 0.8 \pm 0.6$. 
(Note that this number may change slightly as the cutoff increases and deep, unphysical trimers appear.) 
As a comparison, for $^4$He atoms --- which are close but not quite at unitarity --- we extract 0.95 from the values in Table VIII of Ref.~\cite{Hiyama:2011ge}. The corresponding universal number for the first-excited state, $\rho_1$, is harder to nail down by direct calculation.

At NLO, linear corrections are expected from the two-body scattering length and effective range, and from the four-body force. Corrections from two-body currents are expected only at higher orders. Thus one can write relations similar to Eqs. \eqref{total0} and \eqref{total1} for $\langle r^2 \rangle_{4,0}^{(1)}$ and $\langle r^2 \rangle_{4,1}^{(1)}$ multiplied by their respective LO energy splittings from the three-body ground state. For the excited state in particular one can write, up to higher-order terms,
\begin{equation}
m\left(B_{4,1}^{(0)}-B_{3}\right)\text{Re}\langle r^2 \rangle_{4,1}^{(1)} 
= \tilde\rho_1
-\frac{\rho_1}{\kappa_1-1}
\left(\frac{B_{4,1}^{(1)}}{B_{3}} - \kappa_1 \right) ,
\label{r411}
\end{equation}
where $B_{4,1}^{(1)}/B_{3}$ is given by Eq. \eqref{totalexcited} and 
\begin{equation}
\tilde\rho_1(B_3,B_{4,0}^{(1)},a_2,r_2)
\equiv m\left(B_{4,1}^{(1)}-B_{3}\right) \text{Re}\langle r^2 \rangle_{4,1}^{(1)}.
\label{r4110}  
\end{equation}
Potential-model calculations for $^4$He atoms typically include all orders in the deviation from unitarity but, to the extent that higher-order corrections are small, they allow for an estimate of $\tilde\rho_1$. The corresponding number for the LM2M2 potential can again be obtained from Ref.~\cite{Hiyama:2011ge} as 0.0851. The value of $\langle r_{iG}^2 \rangle$ is not given in Ref.~\cite{Lazauskas:2006}, but using its value for $\langle r_{ij}^2 \rangle$ and Eq. \eqref{ric_rij_234} we obtain 0.0406 for $\tilde{\rho}_1$ with the same potential. In Ref. \cite{Lazauskas:2006}, an approximation for the tetramer excited-state wavefunction was made, which might be the reason for the discrepancy with Ref. \cite{Hiyama:2011ge}.

Since the four-body excited state is very close to the 3+1 break-up threshold, four-body calculations are made difficult by the large distances involved. These and other correlations can be obtained, however, from an additional expansion that exploits the separation between the characteristic scale of the trimer and that of the excited tetramer. We describe this approach in the next section.

\section{The Excited State as a Halo}
\label{sec:halo}

Because the four-body excited state is so close to the 3+1 break-up threshold, it must be a halo state where one particle orbits at a large distance from a three-particle core. Such type of system can be described by a low-energy EFT, Halo EFT \cite{Bertulani:2002sz}, where the three-body subsystem is treated as an ``elementary'' particle. This EFT is not sensitive to details of the trimer structure, where the characteristic momentum for each particle is $M_{\rm hi}^{\rm halo}\sim \sqrt{2m B_3/3}$. (As discussed below, the mean-square radius of the unitary three-body cluster does not provide a more stringent estimate.) The four-boson excited state is then effectively a two-body bound state with a characteristic momentum $M_{\rm lo}^{\rm halo}\sim \sqrt{2\mu (B_{4,1}-B_{3})}$, where $\mu\simeq 3m/4$ is the reduced mass. Observables related to this state can be obtained in an expansion in the small ratio $M_{\rm lo}^{\rm halo}/M_{\rm hi}^{\rm halo}\sim 3\sqrt{\kappa_1-1}/2\simeq 1/14$.

Note that the ground tetramer involves momenta $\sim \sqrt{2\mu (B_{4,0}-B_{3})} \sim \sqrt{(\kappa_0-1)2\mu B_3} >M_{\rm hi}^{\rm halo}$ and thus cannot be described in this Halo EFT. Since it involves even larger momenta $\sim 22.7 \sqrt{2\mu B_3}\gg M_{\rm hi}^{\rm halo}$, the decay of an excited tetramer into a lower trimer must be described in Halo EFT by imaginary parameters. The interaction strengths must be complex and translate into complex threshold parameters like the scattering length and the effective range, respectively
\begin{eqnarray}
    a_0&=& a_r + i a_i,
    \\
    r_0&=& r_r + i r_i.
\end{eqnarray}
While $a_r^{-1}={\cal O} (M_{\rm lo}^{\rm halo})$ reflects the shallowness of the excited state, the effective range carries information about the trimer structure and we expect $r_r^{-1}={\cal O} (M_{\rm hi}^{\rm halo})$. The real parts of the shape parameters are expected to scale with $M_{\rm hi}^{\rm halo}$ similarly to the effective range, that is, according to their dimensions. Since the imaginary parts are driven by physics outside the EFT, they should be suppressed compared to the real parts by at least one power of the breakdown scale, for example $|a_i^{-1}|={\cal O} (M_{\rm hi}^{\rm halo})$ and $|r_i^{-1}|={\cal O} ((M_{\rm hi}^{\rm halo})^2/M_{\rm lo}^{\rm halo})$.

Halo EFT can be applied at any order of the Short-Range EFT expansion, where the scattering parameters between halo particle and core can in principle be calculated. As long as the Short-Range EFT expansion converges, the exact values of these parameters will change from order to order but their orders of magnitude, and thus the organization of the Halo EFT expansion, will not. In fact, any underlying theory that yields the same hierarchy of scales for the four-body excited state is amenable to Halo EFT. We start with LO in Short-Range EFT, using the equivalent results of Ref.~\cite{Deltuva:2010xd}, where the effects of the finite range in a potential model were minimized by studying successively higher trimer excited states labeled $n=1,\ldots$. Around $n=5$ properties of the associated excited tetramer vary little and offer a good estimate of the unitarity limit, 
\begin{eqnarray}
    a_r^{(0)-1}/\sqrt{2\mu (B_{4,1}^{(0)}-B_{3})} &\simeq& 0.92, 
    \qquad -a_i^{(0)-1}/a_r^{(0)-1} \simeq 20.7
    \nonumber
    \\
    r_r^{(0)-1}/a_r^{(0)-1} &\simeq& 7.02, 
    \qquad -r_i^{(0)-1}/a_r^{(0)-1} \simeq 36^2.
    \label{EREparam}
\end{eqnarray}
Both $|a_i^{(0)}|/a_r^{(0)}$ and $r_r^{(0)}/a_r^{(0)}$ are within a factor 2 of the expected expansion parameter. Even though $|r_i^{(0)}|/a_r^{(0)}$ is somewhat smaller than expected, in the following we conservatively treat it as if had the expected size. Ratios of similar size are obtained for the lower ($n= 1, 2, 3, 4$) tetramer excited states \cite{Deltuva:2010xd}. Note that we continue to use the superscript $^{(0)}$ to denote the values of parameters from Short-Range EFT at LO, which represents the unitarity limit. For notational simplicity we do not add an index to make explicit the order in Halo EFT.

With this scaling of parameters, Halo EFT for the excited tetramer is similar to Short-Range EFT, just with the additional imaginary parameters at higher orders. The LO, nonderivative two-body contact interaction has a real strength determined by this state's binding energy, $B_{4,1}-B_{3}$ --- or, alternatively, by the real part of the scattering length, $a_r$, the difference being compensated at higher orders. At NLO, the same interaction has an imaginary strength determined by $a_i$, in addition to a two-derivative interaction determined by the real part of the effective range, $r_r$. At N$^2$LO, the latter gets an imaginary strength fixed by $r_i$. Shape parameters, as well as partial waves higher than $S$, start contributing at N$^3$LO. 
The corresponding Hamiltonian is therefore Hermitian at LO and becomes non-Hermitian at NLO. 
The non-Hermitian Hamiltonian at NLO is a consequence of reducing the four-body tetramer excited state dynamics to an effective two-body problem in Halo EFT --- deeper trimers, to which the excited tetramer could decay to, are integrated out of the formalism, and therefore tetramer instability ought to be taken into account by an imaginary term in the two-body effective Hamiltonian. 

\begin{figure}[t]
    \centering
    \includegraphics[width= 15cm]{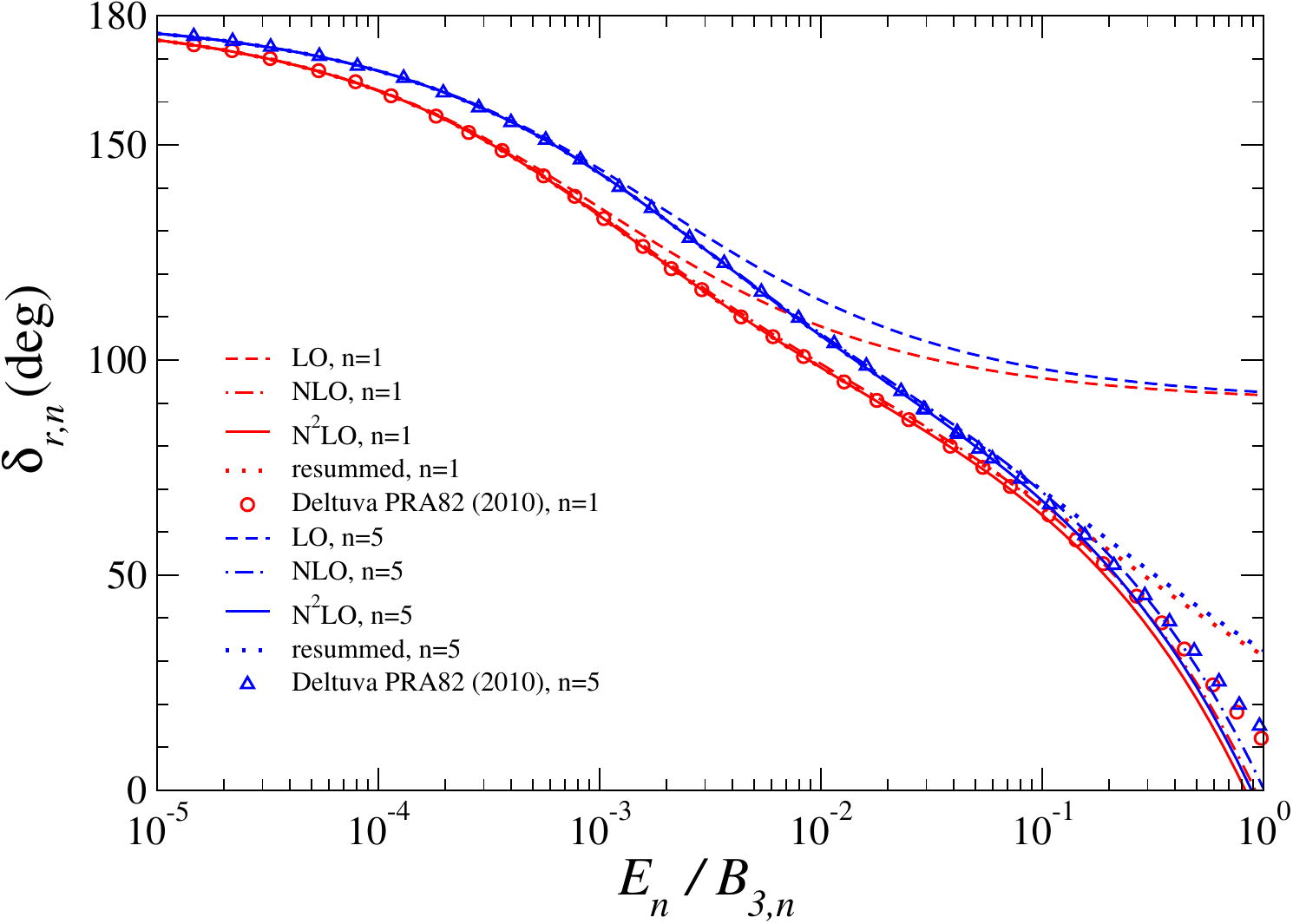}
    \caption{The $S$-wave phase shifts (in degrees) for particle scattering from the $n$th-excited trimer, $\delta_{r,n}$, as functions of the energy $E_n$ relative to the trimer energy $B_{3,n}$. Results from Halo EFT at LO (dashed lines: red for $n=1$, blue for $n=5$), NLO (dash-dotted lines: red for $n=1$, blue for $n=5$), and N$^2$LO (solid lines: red for $n=1$, blue for $n=5$) are compared with the direct results for $n=1$ (red circles) and $n=5$ (blue triangles) from Ref. \cite{Deltuva:2010xd}. Dotted lines represent the corresponding results when the effective-range effect is resummed.
    }
    \label{fig:ps_halo_complex_EREparam}
\end{figure}

\begin{figure}[t]
    \centering
    \includegraphics[width=15cm]{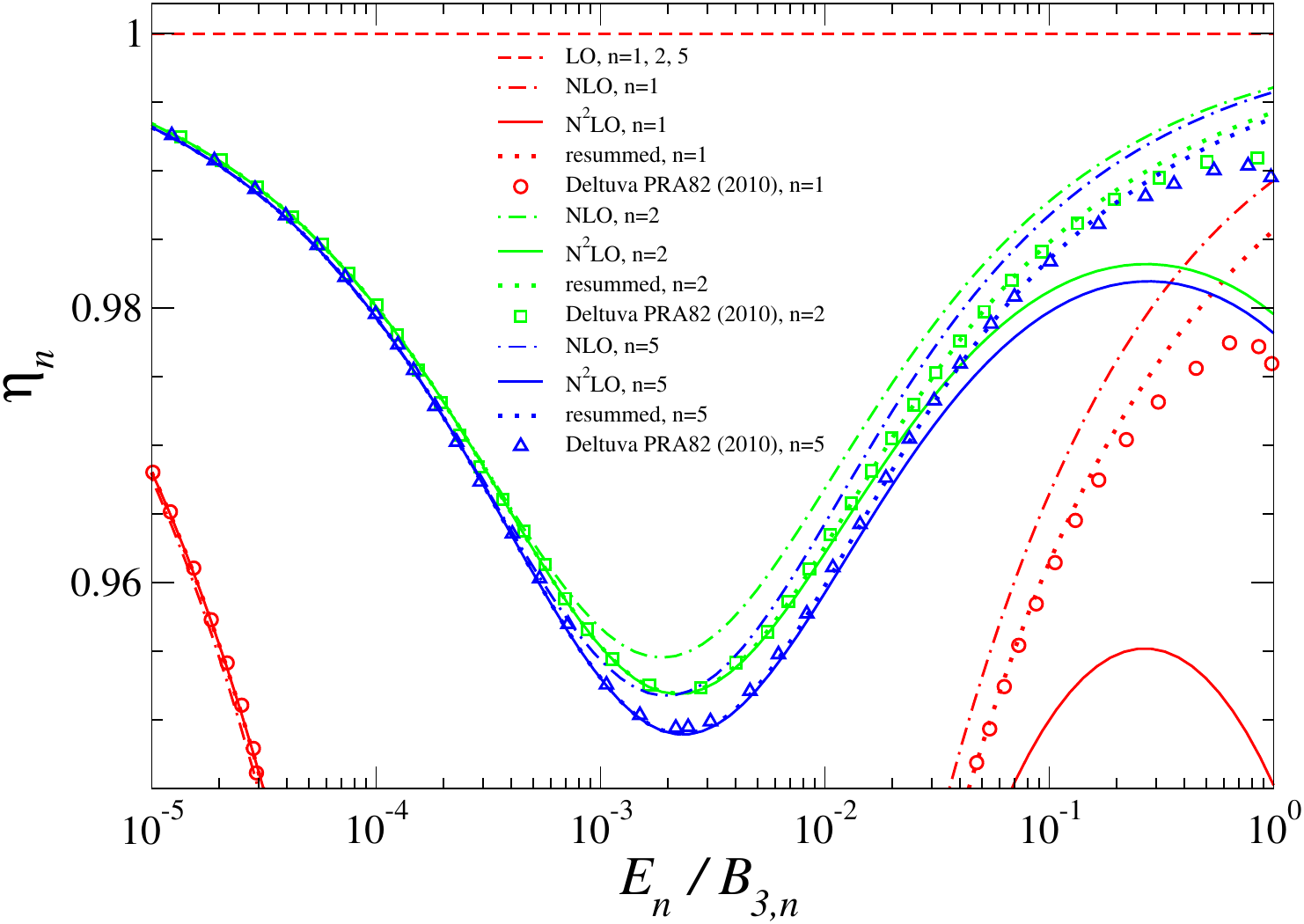}
    \caption{The $S$-wave inelasticity for particle scattering from the $n$th-excited trimer, $\eta_n$, as a function of the energy $E_n$ relative to the trimer energy $B_{3,n}$. Red and blue curves and points as in Fig. \ref{fig:ps_halo_complex_EREparam}. Green curves and square points represent the corresponding results for the $n=2$ tetramer excited state. The LO result for the inelasticity is the same ($\eta_n=1$) for all states.
    }
    \label{fig:inel_halo_complex_EREparam}
\end{figure}

To account for inelasticities, the $S$ matrix in the $S$ wave at energy $E\equiv k^2/2\mu$ can be written as
\begin{equation}
    S(k)=\eta(k) \exp (2i\delta_r(k))
    \equiv \exp (2i\delta (k))
\end{equation}
in terms of real phase shift $\delta_r$ and the inelasticity $\eta$. In the two-body scattering we are considering here, the EFT reduces to an ordering of the effective-range expansion,
\begin{equation}
    k \cot \delta(k)= -\frac{1}{a_r} + i\frac{a_i}{a_r^2} + \frac{r_r}{2}k^2
    + \frac{a_i^2}{a_r^3} + i\frac{r_i}{2}k^2 + \ldots.
    \label{haloERE}
\end{equation}
For the first three orders in the Halo EFT expansion,
\begin{eqnarray}
    \delta_r(k) &=& \pi-\arctan (ka_r) - \frac{r_r a_r^2 k^3}{2(1+k^2 a_r^2)} 
    - \left(\frac{a_i^2}{a_r^2} + \frac{r_r^2k^2}{4}\right) \frac{a_r^3 k^3} {(1+k^2 a_r^2)^2}
    + \ldots,
    \\
    \eta(k) &=& 1+ \frac{2 a_i k}{1+k^2 a_r^2} 
    + \frac{2a_i a_r k^2}{(1+k^2 a_r^2)^2} 
    \left(\frac{a_i}{a_r}+r_r k\right) 
    + \frac{r_i a_r^2 k^3}{1+k^2 a_r^2} 
    +\ldots.
\end{eqnarray}
In Figs. \ref{fig:ps_halo_complex_EREparam} and \ref{fig:inel_halo_complex_EREparam}, respectively, the real phase shift $\delta_r$ and the inelasticity $\eta$ from Halo EFT are plotted as functions of the energy, using as input the unitary values of scattering parameters in Eq. \eqref{EREparam} \cite{Deltuva:2010xd}. In both figures we show results obtained from the $n=5$ excited trimer, but also for $n=1,2$, where we use instead of Eq. \eqref{EREparam} the corresponding values in Ref. \cite{Deltuva:2010xd}. In the energy region we plot, the expansion in Halo EFT displays a very good convergence pattern. Moreover, Halo EFT results converge to the phase shift and inelasticity calculated directly from Ref. \cite{Deltuva:2010xd} for $E\lesssim 0.1 B_3$, a bit lower than the estimated breakdown at $\sim 4B_3/9$. Resumming the imaginary part of the scattering length and the effective range --- that is, treating them exactly instead of expanding as in Eq. \eqref{haloERE} --- gives more visible improvement for the inelasticity than for phase shifts. This is not surprising given that $\eta-1$ starts at NLO and is, consequently, of smaller magnitude --- note the scale of the vertical axis in Fig. \ref{fig:inel_halo_complex_EREparam}. 

We can now translate this scattering input into properties of the excited state.

\subsection{Energy}

The $S$ matrix has a complex pole in the complex $k$ plane at $k_r+ik_i$ with 
\begin{eqnarray}
    k_i&=& \frac{1}{a_r}\left(1+ \frac{r_r}{2a_r}-\frac{a_i^2}{a_r^2}+ \frac{r_r^2}{2a_r^2}+\ldots\right),
    \\
    k_r&=& \frac{a_i}{a_r^2}\left(1+ \frac{r_r}{a_r}-\frac{r_i}{2a_i} +\ldots\right).
\end{eqnarray}
It corresponds to a relative binding energy 
\begin{equation}
    \frac{B_{4,1}-B_3}{B_3} \equiv \frac{\kappa^2}{2\mu B_3}
    = \frac{1}{2\mu B_3 a_r^2} 
    \left( 1+ \frac{r_r}{a_r}+\frac{5 r_r^2}{4a_r^2}
    -\frac{3a_i^2}{a_r^2}+ \ldots \right).
    \label{B_halo_complex_ERE}
\end{equation}
However, this bound state is unstable with a relative half-width 
\begin{equation}
    \frac{\Gamma_{4,1}}{2B_3}
    = -\frac{1}{2\mu B_3 a_r^2} \frac{2a_i}{a_r} 
    \left( 1+ \frac{3r_r}{2a_r}-\frac{r_i}{2a_i}
    + \ldots \right), 
\label{Gamma_halo_complex_ERE}
\end{equation}
which is ${\cal O}(M_{\rm lo}^{\rm halo}/M_{\rm hi}^{\rm halo})$ compared to the relative binding energy, that is, an NLO effect in Halo EFT. Note that a positive width requires $a_i<0$, which means that at NLO the bound state is displaced from the imaginary momentum axis to the left half-plane, the generic situation for an unstable bound state \cite{Badalian:1981xj}. Once we plug in the numbers from Eq. \eqref{EREparam}, 
\begin{eqnarray}
    \kappa_1-1= \frac{B_{4,1}^{(0)}}{B_3} -1 &=& (1.958 + 0.279 + 0.0360 +\ldots) \cdot 10^{-3} \simeq 2.27\cdot 10^{-3} + \ldots, 
    \\
    \gamma_1=\frac{\Gamma_{4,1}^{(0)}}{2B_3} &=& (1.89 + 0.39 +\ldots)\cdot 10^{-4} \simeq 2.28 \cdot 10^{-4} + \ldots,
\end{eqnarray}
to be compared to the direct determination from Refs. \cite{Deltuva:2010xd, Deltuva:2011ae}, $\kappa_1\simeq 2.28 \cdot 10^{-3}$ and $\gamma_1 \simeq 2.38 \cdot 10^{-4}$. The convergence pattern suggests an expansion parameter for energies $\sim 15\%$, which is about twice what was expected. 

We have thus shown that the tetramer excited state is a halo system in the sense that it can be described consistently by Halo EFT. This EFT incorporates correlations among excited-tetramer properties. For example, there is a relation between $B_{4,1}$ and $\Gamma_{4,1}^2$ through Eqs. \eqref{B_halo_complex_ERE} and \eqref{Gamma_halo_complex_ERE}. As such, we can estimate through the expansion of Halo EFT some correlation parameters that are not easy to determine directly in Short-Range EFT.

\subsection{Radius}

We illustrate this idea with the correlation parameter $\rho_1$ in Eq. \eqref{r410} up to N$^2$LO in Halo EFT using the complex scattering length and effective range for atom-trimer scattering at the unitarity limit from Ref. \cite{Deltuva:2010xd}. The real part of the energy is expanded in Eq. \eqref{B_halo_complex_ERE}.  As discussed in App. \ref{rho1_HaloEFT}, there are two types of contributions to the mean-square radius $\langle r_{iG}^2 \rangle$: from the average squared distance between the particle and the core, $\langle r_{ic}^2 \rangle$, and from the core size, $\langle r_{\text{core}}^2 \rangle$.

The expansion for the average distance can be adapted from Ref. \cite{Wu:2022} using the formalism of non-Hermitian quantum mechanics \cite{Moiseyev:1998gjp} or through the form factor \cite{Kaplan:1998sz}, to give
\begin{eqnarray}
    \langle r_{ic}^2 \rangle =\frac{a_r^2}{2}
    \left( 1 + 2i \frac{a_i}{a_r}-\frac{a_i^2}{a_r^2} + \frac{r_r^2}{4a_r^2}+\ldots
 \right).
\end{eqnarray}
At LO, we recover the simple relation 	
\begin{equation}
		\left\langle r_{ic}^2 \right\rangle  = \frac{1}{2 \kappa^2} +\ldots
\label{r2_exp}
\end{equation}
that follows from the LO wavefunction
\begin{equation}
    	\psi(\vec{r}_{ic}) = \sqrt{\frac{ \kappa}{2\pi}} \frac{e^{-\kappa r_{ic}}}{r_{ic}} +\ldots
\end{equation}
At NLO and higher, one has to apply the non-Hermitian formalism of quantum mechanics. The wavefunction obtained from the residue of the Green's function at the unstable bound-state pole is then identified as an eigenstate of the non-Hermitian Hamiltonian. It has, besides the exponential fall off, an oscillatory component, which makes the radius complex in the non-Hermitian formalism.

Apart from the range correction, we should also consider the correction due to the finite size of the core, which is expected to be N$^2$LO. In the unitarity limit, the mean-square radius, like everything else, is determined by $B_3$. In fact, it is known to be given by  \cite{Braaten:2004rn}
\begin{equation}
    \left\langle r_{\text{core}}^2 \right\rangle =\frac{1+s_0^2}{9m B_3},
    \label{mB3r3s}
\end{equation}
in which $s_0 \simeq 1.00624$. Note that $\left\langle r_{\text{core}}^2 \right\rangle^{1/2}$ is approximately 3 times smaller than the smallest distance $\sim (2mB_3/3)^{-1/2}$ Halo EFT is expected to describe.

Putting these results together,
\begin{equation}
    \rho_{1} = \frac{1}{16} \left[ 1+ \frac{r_r}{a_r}
    + \frac{3 r_r^2}{2 a_r^2} 
    -\frac{  4 a_i^2}{ a_r^2}
    + \frac{4}{3} (1+s_0^2) (\kappa_1-1)
    +\ldots \right].
    \label{rho1_halo}
\end{equation}
Using the values from Ref. \cite{Deltuva:2010xd}, one obtains
\begin{equation}
    \rho_{1} = \frac{1}{16} \left( 1+ 0.1425 + 0.0273 + \ldots \right)
    \simeq 0.0731.
    \label{rho1_halo_value}
\end{equation}
The ratio of the various orders is close to the 15\% inherited from the real part of the effective range. While entering at N$^2$LO because $\kappa_1-1={\cal O}((M_{\rm lo}^{\rm halo})^2/(M_{\rm hi}^{\rm halo})^2)$, the finite-core correction is only about 20\% of the N$^2$LO total, which is dominated by effective-range effects.

Equation \eqref{rho1_halo} gives a universal prediction for unstable excited states in the four-body tower from Halo EFT because we use the universal numbers at the unitarity limit. The first term in brackets of Eq. \eqref{rho1_halo} is the LO result. It is parameter-free and therefore holds also away from unitarity --- it is purely a consequence of the halo character of the excited tetramer. Since for the excited state under consideration 
\begin{equation}
		\frac{\kappa^2}{2 \mu} = B_{4,1}^{(n)}-B_3 ,
\end{equation}
at order $n$ in Short-Range EFT, we arrive at 
\begin{equation}
 m (B_{4,1}^{(n)}-B_3) \,\text{Re}\langle r^2 \rangle^{(n)}
 = \frac{1}{16} = 0.0625
 \label{rho_1_halo_LO}
\end{equation}
in LO in Halo EFT. In particular, at NLO in Short-Range EFT,
\begin{equation}
 \tilde \rho_1 = \rho_1 = \frac{1}{16}. 
 \label{tilderho_1_halo_LO}
\end{equation}
Differences appear from NLO on in Halo EFT, where parameters arise which depend on the deviation from the unitarity limit. The scattering length $a_r$, which is very sensitive to the proximity to the trimer threshold, could change significantly, but other ERE parameters should vary less, so we might well expect corrections to Eq. \eqref{tilderho_1_halo_LO} in the same range of 10-30\% as seen above for energies.

As a comparison, our LO Halo EFT value \eqref{tilderho_1_halo_LO} differs by $\simeq$ 26.5\% from the value for the $^4$He tetramer excited state with the LM2M2 potential from Ref. \cite{Hiyama:2011ge}, and lies in-between the values of this reference and the value extracted from Ref. \cite{Lazauskas:2006}. We can estimate the changes due to the effective range and the size of the core,
\begin{equation}
    \tilde{\rho}_1 \simeq m (B_{4,1}-B_3) 
    \, \text{Re} \langle r_{iG}^2 \rangle 
    =\frac{1}{16} \left[1+ 
    \frac{r_0}{a_0}+ \frac{3 r_0^2}{2a_0^2} +\ldots
    + 12 m \left(B_{4,1}-B_3\right) \langle r_{\text{core}}^2 \rangle\right]
    \label{Delta_rho1_range+core}
\end{equation}
in terms of the (real) scattering length $a_0$ and effective range $r_0$ for atom-trimer scattering. Taking the values for $^4$He from Ref. \cite{Lazauskas:2006}, $a_0\simeq 103.7$ \AA, $r_0\simeq 29.1$ \AA, $m\simeq 0.0825$ K$^{-1}$\AA$^{-2}$, $B_{4,1}-B_3\simeq 1.11$ mK, and $\langle r_{\text{core}}^2 \rangle\simeq 39.97$ \AA$^2$ (converted from the mean-square interparticle distance using the relation for $N=3$ in Eq. (\ref{ric_rij_234})),
\begin{equation}
    \tilde{\rho}_1 
    = 0.0625 + 0.0175 + 0.0074 + \ldots + 0.0027 
    \simeq 0.0901.
\end{equation}
This number is only $\simeq 6\%$ away from the result 0.0851 of Ref. \cite{Hiyama:2011ge}. Using instead the core values from Ref. \cite{Hiyama:2011ge}, the core contribution decreases slightly to 0.0023.

\section{Conclusion}
\label{sec:Conc}

Among the many surprising features of Efimov physics, 
one of the most mysterious is the emergence of two towers of geometrically spaced states in the four-boson system. The second tower is placed so very close below to the geometric tower in the three-boson system that normally one would be tempted to invoke fine tuning. Yet, the splitting between towers is an intrinsic property of the dynamics tied to discrete scale invariance. Whatever its explanation,
the proximity to the three-body threshold means that the properties of the states in this excited four-body tower are vulnerable to interactions that break DSI.

In this work we used effective field theory methods to exploit this proximity and reach several conclusions about the properties of the excited four-body state:

\begin{itemize}

\item 
We have pointed out that the organization of interactions intrinsic to EFT leads to definite correlations among four-body observables.
The EFT at LO automatically leads to DSI and gives rise to simple correlations that are partially known in the literature. NLO interactions modify correlations that exist at LO. In addition to a finite two-body scattering length and effective range, renormalization {\it requires} a four-body parameter.

\item 
We have used arguments based on (distorted-wave) perturbation theory to derive the explicit form of the correlation between first-excited and ground states of the two four-body towers at NLO in Short-Range EFT, Eq. \eqref{totalexcited}.
The validity of this form relies on the perturbativeness of the four-body force and, therefore, is a signal of the importance of the three-body scale relative to the four-body scale.

\item 
The large size of the excited state makes a direct calculation of the correlation parameters challenging.
We have used a variety of existing model calculations to constrain some of the parameters appearing in the correlation between excited and ground four-body energies.

\item 
We found consistency among various model calculations, which allows them to be interpreted through Short-Range EFT. In particular, we found that the results from Short-Range EFT where a four-body scale enters perturbatively at NLO are reasonably close to {\it part} of the energy regime explored in the calculations of Refs. \cite{Hadizadeh:2011qj,Hadizadeh:2011ad}, where a four-body scale enters on the same footing as the three-body scale. Thus we conclude that within that limited energy regime the four-body scale is really a subleading effect.

\item 
The fact that part of the results from Refs.~\cite{Hadizadeh:2011qj,Hadizadeh:2011ad} cannot be described by the correlation \eqref{totalexcited} leads to the conjecture that the calculations of Refs.~\cite{Hadizadeh:2011qj,Hadizadeh:2011ad} represent a different universality class of systems where a four-body force enters at LO --- see, e.g., Refs.~\cite{Frederico:2019bnm,Frederico:2023fee}. 

\item
We have pointed out that correlations between excited and ground states exist involving other properties of these states. The proximity between excited four-body and three-body energies is reflected in the large size of these four-body states --- a size much larger than of the three-body state. We have in particular derived the NLO correlation, Eq. \eqref{r411}, between the radius and energy of the excited state.

\item
We have shown how a different EFT, Halo EFT, yields correlation parameters involving the excited state's size. Halo EFT treats a three-body Efimov state as a single unit, reducing the excited state to a two-body problem. We have constructed the theory and its organizational principle.

\item We demonstrated the usefulness of this approach by first exhibiting its consistency with existing calculations of scattering and bound-state properties. Next, we determined parameters appearing in correlations between the radius and energy of the four-body excited states at unitarity, and showed they are close to those of existing calculations away from unitarity. In other words, we have reproduced the result of a difficult four-body calculation thanks to a correlation based on EFT power counting.

\end{itemize}

While we have not explained the emergence of the four-body excited tower at unitarity, we have laid the foundation for the understanding of how its properties change with a small explicit breaking of discrete scale invariance. 
The limited validity of Eq. \eqref{totalexcited} raises the interesting question of whether subleading effects in Short-Range EFT can invert the relative ordering between the four-body excited state and the associated three-body state. Without a full NLO calculation in Short-Range EFT for the excited state, we cannot offer a conclusive answer. For one, we were unable to quantify from the literature the effect of the two-body effective range captured by the correlation parameter $z$. The results given in Fig. \ref{fig:HamPlat} indicate that the two-body scattering length cannot drive the instability while NLO is perturbative and the relative change in $B_{4,0}/B_3$ remains small. By the same token, Fig.~\ref{fig:Delt} suggests that a moderately repulsive four-body force might lead to instability. Even in this case, it is not clear that level inversion can be captured without a rearrangement of the Short-Range EFT expansion. Examples of a need for change in EFT power counting in specific energy regimes exist for simpler systems \cite{Hammer:2019poc}, e.g. pion-nucleon scattering in the vicinity of the Delta peak (when the Delta self-energy is no longer a higher-order effect) and Compton scattering on the deuteron at very small energies (when the interaction between two nucleons cannot be neglected in states between photon absorption and emission).

In the future, we intend to address this question with a full four-body calculation to NLO in Short-Range EFT. We would also like to apply these ideas to the interesting changes these states experience \cite{Deltuva:2011ae,Deltuva:2012ms} when the two-body system has a finite scattering length which must be accounted for at LO in Short-Range EFT.

\section*{Acknowledgments}
We thank Arnoldas Deltuva for a communication on his results.
Gratefully acknowledged is the hospitality extended to UvK at USP and ITA and to TF at IJCLab, which was made possible by the CAPES-COFECUB agreement Proj. no. 88887.370819/2019-00 (CAPES) and Projet N$^{\rm o}$ 45050TH (COFECUB). This work was supported in part
by the U.S. Department of Energy, Office of Science, Office of Nuclear Physics, under award DE-FG02-04ER41338 (UvK, FW),  INCT-FNA project 464898/2014-5 (TF, RH) and FAPESP Thematic grants 2017/05660-0, 2019/07767-1 (TF, RH), and 2020/04867-2 (RH).

\appendix
\section{Radii in Short-Range EFT}
\label{sec:ric_rij}

In this appendix, we relate $\langle r_{ij}^2 \rangle$, the average squared distance between two particles, to $\langle r_{iG}^2 \rangle$, the average squared distance of a particle from the center of mass, for a system consisting of $N$ identical particles with mass $m$ at positions $\vec{r}_i\;(i=1,\ldots,N$), as we have in Short-Range EFT. In Secs. \ref{sec:othercorr} and \ref{sec:halo} we use this relation to compare results found in the literature. 

The relative position of particle $i$ to the center of mass is
\begin{eqnarray}
    \vec{r}_{iG} \equiv \vec{r}_i-\vec{R}= \frac{1}{N}
    \sum_{j \neq i} \vec{r}_{ij},
\end{eqnarray}
where 
\begin{eqnarray}
    \vec{R}= \frac{1}{N}\sum_{i=1}^N \vec{r}_i
\end{eqnarray}
is the position of the center of mass and
\begin{eqnarray}
    \vec{r}_{ij}\equiv \vec{r}_i -\vec{r}_j.
\end{eqnarray}
denotes the relative position between two particles $i$ and $j$. 

Since the particles are identical, $\langle r_{iG}^2 \rangle$ should be the same for any particle $i$, and $\langle r_{ij}^2 \rangle$ should be the same for any pair of particles. We then have
\begin{eqnarray}
    N \langle r_{iG}^2 \rangle &=& \sum_{i=1}^N \langle r_{iG}^2 \rangle = \frac{1}{N^2} \sum_{i=1}^N \left\langle \left( \sum_{j \neq i} \vec{r}_{ij} \right)^2 \right\rangle = \frac{1}{N^2} \sum_{i=1}^N \left[ (N-1)\langle r_{ij}^2 \rangle + 2 \sum_{j\neq i} \sum_{k>j}  \langle \vec{r}_{ij} \cdot \vec{r}_{ik} \rangle \right] \nonumber\\
    &=& \frac{N-1}{N} \langle r_{ij}^2 \rangle + \frac{2}{N^2} \sum_{i=1}^N \sum_{j\neq i} \sum_{k>j} \langle \vec{r}_{ij} \cdot \vec{r}_{ik} \rangle.
    \label{ric_sqr_1}
\end{eqnarray}
Using
\begin{eqnarray}
    \vec{r}_{ij} \cdot\vec{r}_{ik}+ \vec{r}_{ji} \cdot\vec{r}_{jk} = r_{ij}^2
\end{eqnarray}
leads to
\begin{eqnarray}
    N \langle r_{iG}^2 \rangle = \frac{N-1}{N}  \langle r_{ij}^2 \rangle + \frac{N_t}{N^2} \langle r_{ij}^2 \rangle,
\end{eqnarray}
in which 
\begin{eqnarray}
    N_{t} = N\left[ (N-2)+(N-1)+\cdots+1  \right] = \frac{N(N-1)(N-2)}{2}
\end{eqnarray}
is the total number of terms summed in the second term in the last equality of Eq. (\ref{ric_sqr_1}). We finally obtain
\begin{eqnarray}
    \langle r_{iG}^2 \rangle = \frac{N-1}{2N} \langle r_{ij}^2 \rangle.
    \label{ric_rij}
\end{eqnarray}
Since we did not make any assumption except that the system is composed of $N$ identical particles, this relation is very general. In particular, 
\begin{eqnarray}
    \frac{\langle r_{iG}^2 \rangle}{\langle r_{ij}^2 \rangle} = \left\{
    \begin{aligned}
    & 1/4=0.25,\;N=2\\
    & 1/3\approx 0.33333,\;N=3\\
    & 3/8=0.375,\;N=4
    \end{aligned}
    \label{ric_rij_234}
    \right.
\end{eqnarray}
The result for $N=2$ is intuitively correct. In Ref.~\cite{Hiyama:2011ge}, the ratios are 0.33315 and 0.33330 for the $^4$He trimer ground and excited states, and 0.37467 and 0.37333 for the $^4$He tetramer ground and excited states, respectively. They are all close to the analytical values in Eq. \eqref{ric_rij_234}. The small deviation may be due to approximations or numerical issues.

\section{Radii in Halo EFT}
\label{rho1_HaloEFT}

Here we discuss the contributions to $\langle r_{iG}^2 \rangle$ in Halo EFT that enter up to, and including, N$^2$LO \cite{Ryberg:2019cvj}. Operators involving two bodies and an external probe, which would bring contributions beyond those from the wavefunction, appear at N$^3$LO \cite{Chen:1999tn}. Thus, if we denote by $f(\vec{\rho})$ the normalized mass distribution, then
\begin{equation}
    \langle r_{iG}^2 \rangle 
    = \left\langle \int d^3 \vec{\rho} \;\rho^2 f(\vec{\rho}) \right\rangle ,
    \label{ric2moregeneral}
\end{equation}
where the average is taken with respect to the wavefunction in Halo EFT.

In Halo EFT at LO and NLO, we consider the first tetramer excited state as two point particles of mass $m_c$ and $m_i$ at positions $\vec{r}_c$ and $\vec{r}_i$ in the center-of-mass frame. The relative coordinate is 
\begin{equation}
     \vec{r}_{ic} \equiv \vec{r}_i -\vec{r}_c = \frac{m_i}{\mu} \, \vec{r}_i 
     = -\frac{m_c}{\mu} \, \vec{r}_c
     \label{r_r1_r2}
 \end{equation}
in terms of the reduced mass $\mu=m_i m_c/(m_i+m_c)$, and the normalized mass distribution, 
\begin{equation}
    f(\vec{\rho})=\frac{\mu}{m_c} \delta(\vec{\rho}-\vec{r}_i) 
    +\frac{\mu}{m_i} \delta(\vec{\rho}-\vec{r}_c) +\ldots
\end{equation}
The dominant contributions to $\langle r_{iG}^2 \rangle$ arise simply from the average of the squared distance $\langle r_{ic}^2 \rangle$.
At N$^2$LO the matter distribution gets distorted by the finite size of the constituents, with the largest contribution from the core radius, $\langle r_{\text{core}}^2 \rangle$. Up to and including this order,
\begin{equation}
    \langle r_{iG}^2 \rangle 
    = \left\langle \int d^3 \vec{\rho} \;\rho^2 f(\vec{\rho})  \right\rangle
    = \frac{\mu^2}{m_i m_c}\langle r_{ic}^2 \rangle 
    + \frac{\mu}{m_i}\langle r_{\text{core}}^2 \rangle +\ldots
    \label{ric2general}
\end{equation}
For a single-particle core, $m_i=m_c$, as considered in Ref. \cite{Wu:2022}, the core size can be neglected and $\langle r_{iG}^2 \rangle\simeq \langle r_{ic}^2 \rangle/4$, as one would expect from a radius. Instead, here we have a core consisting of three bosons and a halo boson. In lowest orders we can neglect the binding energy of the core and  $m_c\simeq 3m_i\equiv 3m$ so that $\mu \simeq 3m/4$. Equation \eqref{ric2general} becomes
\begin{equation}
     \langle r_{iG}^2 \rangle 
     = \frac{3}{4} 
     \left[\left\langle \left( \frac{r_{ic}}{4}\right)^2 \right\rangle 
     + \langle r_{\text{core}}^2 \rangle\right]
     + \frac{1}{4} \left\langle \left( \frac{3r_{ic}}{4}\right)^2 \right\rangle 
     +\ldots
     = \frac{3}{16} \langle r_{ic}^2 \rangle 
     + \frac{3}{4}\langle r_{\text{core}}^2 \rangle +\ldots,
\end{equation}
which is consistent with the fact that a boson is a part of the core with probability $3/4$ and is the halo particle with probability $1/4$.

\end{document}